\documentclass[runningheads]{llncs}
\usepackage[T1]{fontenc}
\usepackage[hyphens]{url}
\usepackage[hidelinks]{hyperref}
\hypersetup{
    colorlinks=false,
    pdfborder={0 0 0}
}
\usepackage{graphicx}
\usepackage{amsmath}
\usepackage[labelfont={bf,it}]{caption}

\begin{document}
\title{Optimal Time Window and Frequency Bandwidth Parameter Combination for Subject-Specific Motor Imagery EEG Classification}
\titlerunning{Optimal Time Window and Frequency Bandwidth for MI Classification}
%
\author{Matthew A. McCartney\inst{[0009-0004-2702-9425]}\inst{1} \& Liisa A. Kivioja\inst{[0009-0003-5248-2566]}\inst{1}, Sonal S. Baberwal\inst{[0000-0002-4809-1568]}\inst{1}
\and
Shirley Coyle\inst{[0000-0003-0493-8963]}\inst{1,2} 
}
\authorrunning{Matthew A. McCartney \& Liisa A. Kivioja et. al }
%
\institute{School of Electronic Engineering, Dublin
City University, D09Y074 Dublin, Ireland \and Insight SFI Centre for Data Analytics and
School of Computing, Dublin City University, D09Y074
Dublin, Ireland.
\email{ matthew.mccartney2@mail.dcu.ie,}\\
\email{liisa.kivioja2@mail.dcu.ie}}
\maketitle              
\begin{abstract}
Motor-imagery (MI) EEG can be classified using supervised machine learning techniques such as Linear Discriminant Analysis applied to features extracted by Common Spatial Patterns.
Performance of these models varies widely, possibly due to MI studies commonly utilising differing post-cue time windows and frequency bands to one another. 
This study aims to assess how the simultaneous optimisation of both these parameters impact MI classification performance. 
This is done by iteratively training and testing a series of subject-specific models on different combinations of frequency bandwidth and time window options across 109 subjects.
This is followed by a statistical analysis using repeated measures ANOVA to uncover significant differences between different bandwidths and time windows in terms of accuracy across the patient cohort. 
The resulting visualisations and statistical tests show that there are, indeed, significant differences between both specific time windows and specific bandwidths in terms of accuracy.
While the comparison of classification accuracies across 23 frequency bandwidths during five different time windows demonstrates an optimal temporal and spectral scale combination of (0, 4) s at the range of (4, 12) Hz across all subjects, the subjects demonstrate similar accuracies for other parameter combinations.
These findings highlight the efficacy of personalised models to detect optimal temporal and spectral parameter combinations to best classify MI EEG signals that inherently vary across subjects.

\end{abstract}
\section{Introduction}
Leveraging brain activity during imagined movement enables the operation of brain-computer interfaces (BCIs) \cite{Pfurtscheller1997}. Imagined movement, known as motor imagery (MI), produces oscillations, mostly alpha/mu (8-13 Hz) and beta (14-30 Hz) waves, in the sensorimotor area. These oscillations can provide information about intended movement in the form of contralateral brain activity in the sensorimotor cortex. For instance, upon performing a left-hand MI task, a BCI system can interpret oscillatory changes within the right brain hemisphere and classify these as a left-hand imagery \cite{Pfurtscheller1999}. Machine learning (ML) classification is pivotal for decoding such changes within the sensorimotor cortex, providing information about intended movement with varying confidence levels \cite{Lotte2014}. 

In particular, extensive research has been carried out using electroencephalography (EEG), to capture sensorimotor signals during MI-BCI protocols e.g. \cite{Zapala2020}; \cite{rimbert2023embc}. EEG is widely used as a non-invasive and low-cost approach to neural signal acquisition, due to its high-temporal resolution, accessibility and ease of use. The capturing of EEG signals requires both temporal parameters, specifying the start and duration of collected post-cue EEG signals, and spectral parameters, specifying the frequency bandwidth of the EEG signal to be acquired \cite{pfurtscheller2006eeg}. 

Research shows that the variation of either temporal parameters e.g. \cite{Tidare2021} or spatial parameters e.g. \cite{Meng2023}; \cite{Zhang2015} in isolation, affects the performance of MI classification models. Such studies have examined the optimal time window and the optimal frequency bandwidth for MI classification individually, but to the authors knowledge, the combinatory effect of these parameters on a MI EEG classification protocol has not been previously studied. As of now, there is no consensus on the optimal combination of temporal and spectral parameters used in EEG based MI-BCI training protocols. This study aims to elucidate the optimal combinations of these EEG parameters for MI classification performance on subject-specific models using a grid search approach. 

Past studies have utilised a wide selection of time windows according to specific experimental set-up designs. These temporal subsets, explored in previous works, vary in duration and start time post cue, potentially resulting in a range of classification performance. The selection of time window options for the current study include: 0 - 2.25 s \cite{rimbert2023embc}, 0 - 4 s \cite{Zapala2020}, 0.5 - 2.5 s \cite{Meng2023}, 0.5 - 3.5 s \cite{rimbert2023embc} and 1 - 3.5 s \cite{Miao2021}. This study opted to examine two sets of frequency bandwidths, a set derived from commonly used bands from the literature: theta, alpha/mu and beta waves; and a set derived from a sliding window mechanism: a window size of 4 Hz, a step size of 2 Hz and ranging from 4 to 40 Hz. This allowed for the examination of the optimal frequency bandwidth at both well-known, broad frequencies and at narrower, specific frequencies. 

These parameter sets can be used to generate a grid of options, with a ML classification model created for each combination. Each model consists of feature extraction and dimensionality reduction using common spatial patterns (CSP) followed by linear discriminant analysis (LDA) to differentiate between MI classes. With multiple models generated for each parameter combination, the performance of each classifier is averaged across a participant cohort to get a generalised view of that combination's performance in a large population. Statistical tests are utilised to confirm or deny a significant difference between the performance of models trained on different parameter combinations.

In addition to this study’s primary goal of elucidating the optimal parameter combination for EEG-based MI classification, inter-subject variability and combination preferences are explored and addressed by analysing the performance of individual participants. The meaning and interpretation of the optimal combination is hypothesised and discussed. In addition, future directions and potential pitfalls are also considered.

\section{Methodology}
\subsection{Dataset}
This study utilised the PhysioNet \cite{Goldberger2000} EEG Motor Movement/Imagery Dataset, comprising EEG data from 109 participants \cite{Schalk2004}. For the 64-electrode channel BCI2000 system, the international 10-10 setup was used. This dataset was chosen due to the large population and the ability of the 64-electrode system to capture subtle changes in MI patterns. This paper focuses on Task 2 of the dataset, where each participant engaged in 12 experimental runs of the hand movement imagery task. Each experimental run lasted two minutes and consisted of 30 trials (15 right-hand and 15 left-hand) \cite{shuqfa2024increasing}. Participants were instructed to imagine opening and closing either their left or right fist, according to the presented cue (\textbf{\textit{Fig. 1}}). The EEG recordings were taken on experimental runs 4, 8 and 12. The detailed description of the data acquisition procedure and experimental setup can be found in \cite{Schalk2004}. 

\begin{figure}[ht]
\centering
\includegraphics[width=0.9\textwidth]{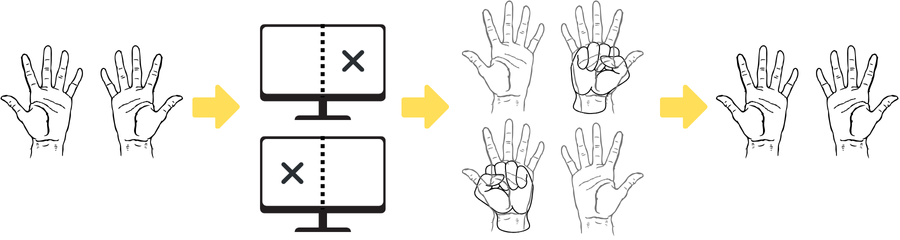}
\caption{\textit{Visualisation of motor imagery “Task 2” protocol used in PhysioNet dataset}}
\label{fig:hands}
\end{figure}

\subsection{Pipeline}

A pre-existing and frequently used pipeline from MNE has been adopted and modified for this study. This pipeline uses CSP for feature extraction followed by an LDA classifier to separate left-hand from right-hand MI EEG signals \cite{MNECSP}. 

\subsection{Feature Extraction}
CSP is a feature extraction technique reminiscent of Principal Component Analysis (PCA). CSP solves a generalised eigenvalue equation between two class covariances while PCA is the eigenvalue decomposition of a single covariance matrix \cite{Parra2003}. CSP creates spatial features that allow for the separation of two classes (e.g. left-hand and right-hand MI) by maximising the variance ratio between the two groups \cite{Antony2022}. In terms of this study, CSP compresses the 64 EEG channels into just four dimensions (CSP1 - CSP4) that maximise the variance between left-hand and right-hand data points. CSP creates an environment where class separation is approximately linearly separable, making simple linear classifiers like LDA well suited for MI decoding \cite{Blankertz2007}.

\subsection{Classification}
Linear Discriminant Analysis or LDA is a supervised learning algorithm used for the linear classification of samples into groups based on discriminant functions \cite{Hayati2024}. These discriminant functions are linear combinations of the original variables. LDA minimises the variance within a group, minimising the scatter within the class, but maximises the variance between groups, maximising the distance between the means of each class \cite{Hidalgo2016}. This highlights the clustering of different groups and allows for the creation of a hyperplane decision boundary between classes \cite{jridi2025explainable}. LDA is often paired with feature extraction, such as CSP in MI research \cite{Lotte2014}.

\subsection{Experiments}
To investigate the relationship between both time window and frequency bandwidth against classification performance, a grid search pipeline was constructed that iterates through various combinations of parameter options for each subject (\textbf{\textit{Fig. 2}}). For each combination, a new set of 4 CSP axes were created and a new LDA classifier was trained and tested on this iteration’s data via 10-fold cross validation, after which the performance was averaged across folds. Shrinkage was later added as a regularisation mechanism, a preventative measure to overfitting. This leaves a grid for each subject, filled with performance scores for every parameter combination. The performance scores of each combination were then averaged across the 109 participants. This gives the average performance of each parameter combination across the subject population.

\begin{figure}[ht]
\centering
\includegraphics[width=0.9\textwidth]{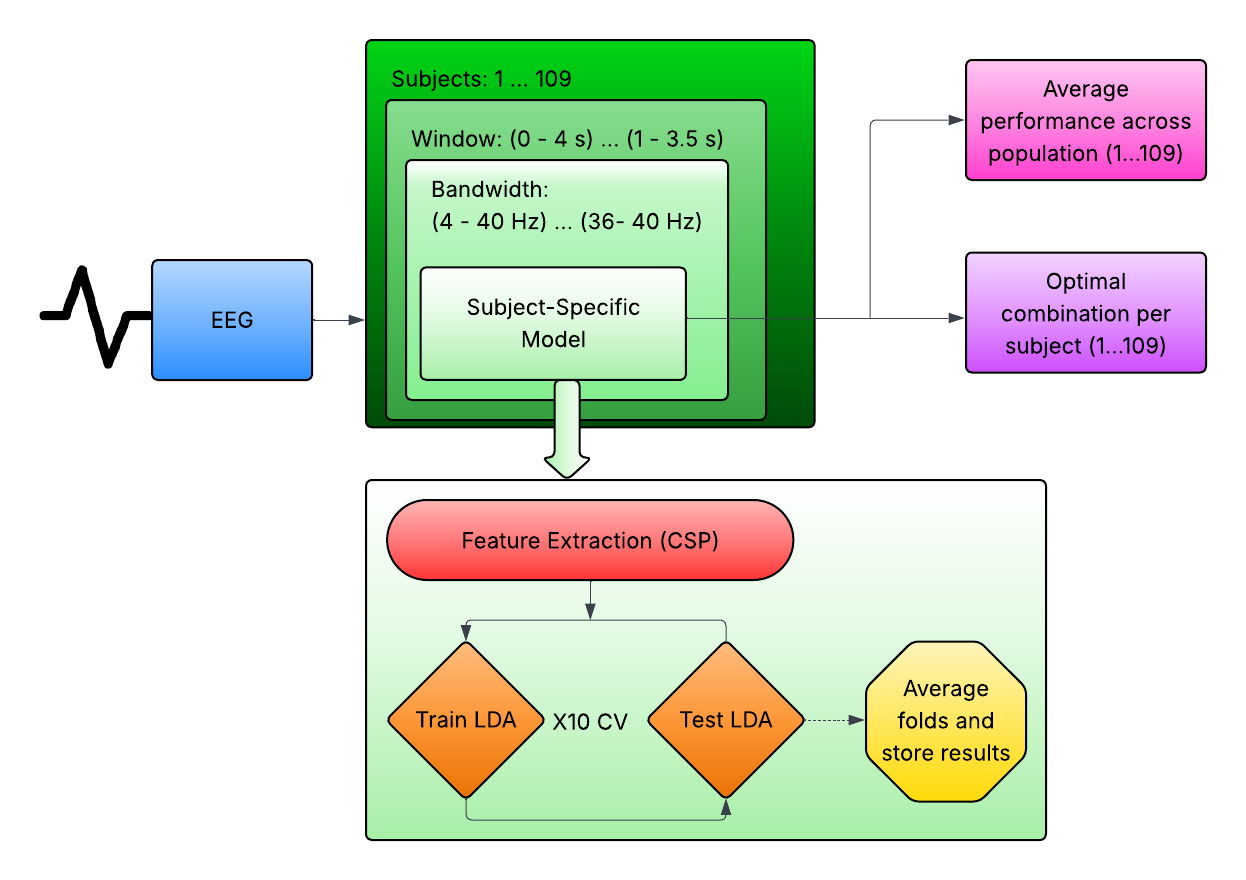}
\caption{\textit{Flowchart of machine learning pipeline iterating through different combinations of bandwidth and windows for each participant: X10 CV - 10 fold Cross Validation.}}
\label{fig:ml_pipeline}
\end{figure}

\subsection{Time Windows and Bandwidth}  
Inspired by a sliding window approach for time windows reported by \cite{gaur2021}, a sliding window approach was used to capture a range of specific frequency bandwidths. The mechanism uses a window size of 4 Hz and a step size of 2 Hz, starting at 4 Hz, slightly preceding the alpha/mu range, and ending at 40 Hz, slightly beyond the beta range into the early gamma frequencies. This gives a gradient of bandwidth frequencies across a wide range, potentially catering for individual variability. 

In addition to the sliding window bandwidth, six other bandwidths that are typically reported in the field of MI were investigated including 4 - 12 Hz (theta and alpha), 8-13 Hz (alpha), 8 - 30 Hz (alpha and beta), 14 - 30 Hz (beta), 14 - 40 Hz (beta and early gamma) and 4 - 40 Hz (broadband) e.g. \cite{Padfield2019}, \cite{rimbert2023embc}, \cite{Zhang2015}.

The time window describes the period of time after the MI cue that should be used to train the classifier. A set of five commonly used options from the literature were chosen. These consisted of  0 - 2.25 seconds \cite{rimbert2023embc}, 0 - 4 seconds \cite{Zapala2020}, 0.5 - 2.5 seconds \cite{Meng2023}, 0.5 - 3.5 seconds \cite{rimbert2023embc} and 1 to 3.5 seconds \cite{Miao2021}, where 0 seconds is the moment the MI cue is presented. This range of window positions and sizes allows for the optimisation of the trade-off between capturing small amounts of relevant data and large amounts of data with some redundancy.

\subsection{Evaluation Metrics and Statistical Tests}
Although accuracy is widely used for MI classification, it is only an appropriate metric for evaluation when two conditions hold: (i) the classes are balanced and (ii) the classifiers’ performance is not biased towards one class \cite{Thomas2013}. 

The current PhysioNet dataset met the first condition as each participant had an approximately equal number of trials for left and right tasks. However, the second condition was not met as class performance frequently differed across participants, with the personalised models favouring either right or left classes. While this was expected due to the well-known preference individuals have to certain hands, the violation of this rule needed to be addressed.

Accuracy was reported as the primary evaluation metric for comparison across combinations and participants but was paired with a secondary, supportive metric, Cohen’s Kappa. Cohen’s Kappa is a metric adjusted for chance agreement that is less sensitive to unbalanced datasets and provides a robust context to each personalised model’s performance. Cohen’s Kappa results were compared against that of accuracy to analyse whether these metrics displayed similar trends. 

Repeated measures ANOVA was used to statistically compare both the accuracy between time window options and the accuracy between frequency band options across participants, following the Bonferroni correction of the p-value.

\section{Results}

\subsection{Combination accuracies averaged across population}
In order to visualise the performance of each combination of frequency band and time window, the average accuracy across participants for each combination was plotted on a heatmap (\textbf{\textit{Fig. 3}}). The optimal time window and frequency band combination across all participants was (0, 4) s at the frequency (4, 12) Hz, while the lowest classification accuracy across all participants was achieved for the combination (1, 3.5) s at the range of (36, 40) Hz. 

\begin{figure}[ht]
\centering
\includegraphics[width=1.0\textwidth]{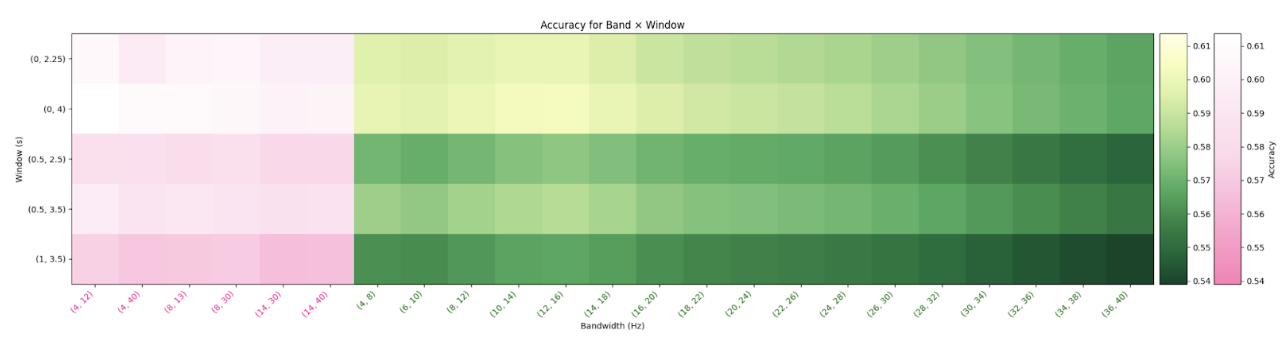}
\caption{\textit{Heatmap of Bandwidth (x-axis) vs Time Window (y-axis) for all combinations averaged across all participants. Pink: Literature derived bandwidths, Green: Sliding Window bandwidths}}
\label{fig:fig3}
\end{figure}

\subsection{Statistical Significance: ANOVA}

The results of every combination for each participant were extracted from the pipeline as CSV files and used for statistical testing.

\begin{table}[ht]
\centering
\caption{Results of the repeated measures ANOVA before the Bonferroni correction.}
\label{tab:anova}
\begin{tabular}{l c}
\hline
\textbf{Effect} & \textbf{p-value} \\
\hline
Frequency Bandwidth & $< 0.001$ \\
Time Window & $< 0.001$ \\
Frequency Bandwidth: Time Window & $0.044$ \\
\hline
\end{tabular}
\end{table}

\begin{figure}[ht]
\centering
\includegraphics[width=0.9\textwidth]{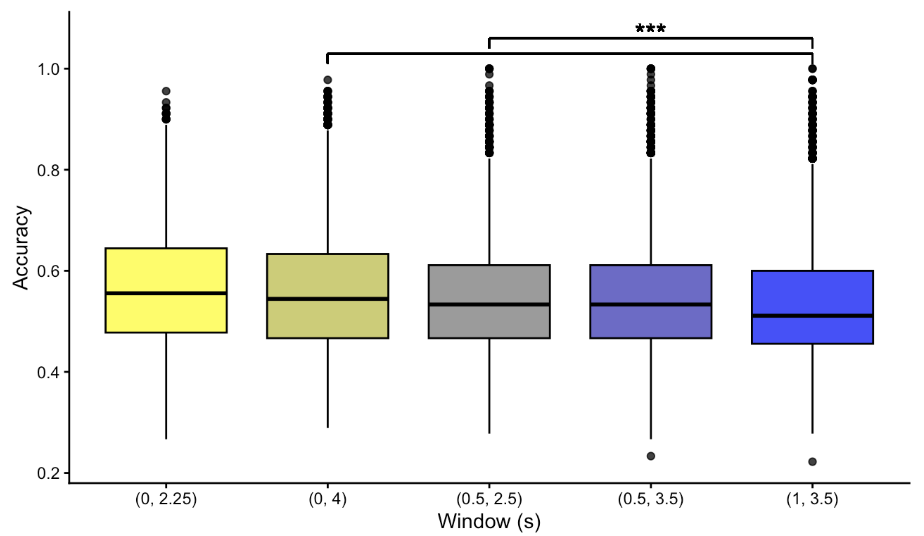}
\caption{\textit{MI EEG classification model accuracies obtained from varying time windows (seconds after cue). Results show significant differences after the Bonferroni correction.}}
\label{fig:fig4}
\end{figure}

The repeated measures ANOVA found statistically significant interactions within and between the time window and frequency bandwidth variables in terms of measured accuracy (\textbf{\textit{Table 1}}). Cohen’s Kappa was also used to depict the statistically significant effects between both frequency bands and time windows, exhibiting similar trends to accuracy (see Supplementary Figures). Following the Bonferroni correction for a representative p-value, calculated to be 0.05/(5 x 23) \text{\(\sim\)} 0.000435, significant interactions of interest were further investigated. 

\subsection{Statistical Significance: Time Windows}
Repeated measures ANOVA following Bonferroni correction revealed significant differences between the time window combinations (0, 4) s with (1, 3.5) s (F(1,108) = 39.20, p < 0.0001), and (0.5, 2.5) s with (1, 3.5) s (F(1, 108) = 38.47, p < 0.0001) (\textbf{\textit{Fig. 5}}). Specifically, the time windows (0, 4) s (M = 0.567, SD = 0.130) and (0.5, 2.5) s (M = 0.547, SD = 0.122) showed higher accuracies, with (0, 4) s time window showing, on average, the highest classification accuracy compared to other time windows. 

\begin{figure}[ht]
\centering
\includegraphics[width=0.9\textwidth]{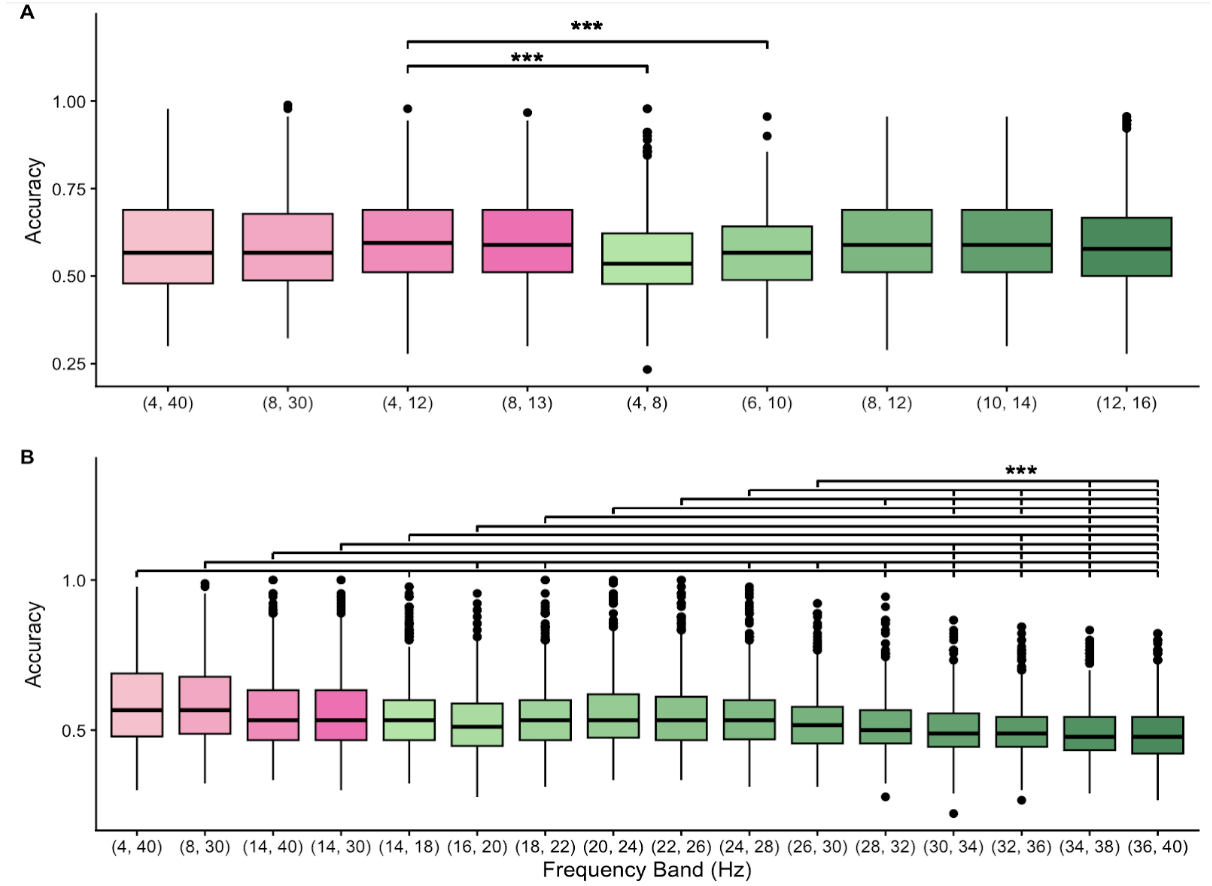}
\caption{\textit{MI EEG classification model accuracies obtained from varying frequency bandwidths across combinations of (A) alpha, alpha and beta, theta-alpha, broadband bandwidths and early alpha-specific bandwidths, and (B) beta, alpha and beta, late-beta, broadband bandwidths and late beta-specific bandwidths. The colour differentiates between broad bandwidth (pink gradient) derived from literature and specific bandwidths (green gradient) derived by sliding window. Results indicated significant differences in both (A) and (B).}}
\label{fig:fig5}
\end{figure}

It is of interest that the time window (0, 2.25) s (M = 0.566, SD = 0.119) showed the second highest average classification accuracy, although the segment had no significant interactions with other time window segments.

\subsection{Statistical Significance: Frequency Bandwidths}
Repeated measures ANOVA following Bonferroni correction indicated significant differences for combinations of broad bandwidths with both alpha- and beta-specific frequencies (\textbf{\textit{Fig. 5A and 5B}}). 

For broad bandwidth combinations with alpha/mu-specific frequencies (\textbf{\textit{Fig. 5A}}), the bandwidth combinations of (4, 12) Hz with (4, 8) Hz (F(1, 108) = 25.81, p < 0.0001), and (4, 12) Hz with (6, 10) Hz (F(1, 108) = 21.48, p < 0.0001) yielded significant differences. Here, the bandwidth range (4, 12) Hz (M = 0.602, SD = 0.132) was significantly higher than both (4, 8) Hz (M = 0.555, SD = 0.116) and (6, 10) Hz (M = 0.568, SD = 0.113) ranges.

For broad bandwidth combinations with beta-specific frequencies (\textbf{\textit{Fig. 5B}}), the broad bandwidths (4, 40) Hz (M = 0.588, SD = 0.139) and (8, 30) Hz (M = 0.592, SD = 0.14) showed higher average accuracies compared to other broad bandwidth ranges. In turn, among the beta-specific frequencies, the bandwidths (20, 24) Hz (M = 0.558, SD = 0.119) and (22, 26) Hz (M = 0.554, SD = 0.118) showed higher average accuracies compared to the other beta-specific frequency ranges. However, there were no significant interactions between these specific combinations of broad bandwidth and beta-specific bandwidth ranges. Therefore, these specified frequency band ranges, although often exhibited with a trend of showing statistically significant differences with other bandwidth ranges around them, showed no statistical differences among each other. 

\subsection{Optimal Parameter Combinations per Participant}
In addition to analysing the average performance of each parameter combination, the optimal parameter combinations for each individual were investigated. The optimal parameter combinations for each participant were ranked by accuracy and the top 10 performing combinations and subjects can be seen in \textbf{\textit{Table 2}} below. 

The highest performance was from participant 29 with an accuracy of 1, indicating that this model was able to classify left-hand from right-hand MI perfectly for this participant. Notably, the participant's optimal parameter combination is a frequency bandwidth of (14, 40) Hz and a time window of (0.5, 2.5) seconds, differing from the results seen for the average population. It is also important to note, Cohen’s Kappa followed the trends seen in accuracy values.

\begin{table}[ht]
\centering
\caption{\textit{Top 10 participants from optimal results ranked by accuracy score}}
\label{tab:top10}
\begin{tabular}{c c c c c c c}
\hline
\textbf{Subject} & \textbf{Band} & \textbf{Window} & \textbf{Accuracy} & \textbf{Acc\_std} & \textbf{Kappa} & \textbf{Kappa\_std} \\
\hline
29  & (14, 40) & (0.5, 2.5) & 1.000000 & 0.000000 & 1.000000 & 0.000000 \\
90  & (4, 8)   & (0, 4)     & 0.977778 & 0.044444 & 0.956098 & 0.087805 \\
69  & (4, 12)  & (1, 3.5)   & 0.977778 & 0.044444 & 0.956098 & 0.087805 \\
54  & (14, 18) & (1, 3.5)   & 0.977778 & 0.044444 & 0.953846 & 0.092308 \\
62  & (8, 13)  & (0.5, 2.5) & 0.966667 & 0.050918 & 0.931895 & 0.104073 \\
94  & (14, 18) & (0, 4)     & 0.955556 & 0.054433 & 0.911069 & 0.108961 \\
40  & (6, 10)  & (0, 2.25)  & 0.955556 & 0.054433 & 0.911069 & 0.108961 \\
34  & (10, 14) & (0.5, 3.5) & 0.955556 & 0.054433 & 0.911069 & 0.108961 \\
7   & (8, 30)  & (0.5, 3.5) & 0.955556 & 0.054433 & 0.912195 & 0.107539 \\
60  & (12, 16) & (0, 4)     & 0.944444 & 0.074536 & 0.884527 & 0.156758 \\
\hline
\end{tabular}
\end{table}

The most frequently occurring optimal time window and frequency band combinations, as seen in \textbf{\textit{Fig. 6}}, are for time windows (0, 2.25) s (N = 34, M = 0.747, SD = 0.086) and (0, 4) s (N = 31, M = 0.79, SD = 0.107). While occurring with the lowest number of optimal combinations, the highest average classification accuracy was achieved in the time window (1, 3.5) s (N = 10, M = 0.817, SD = 0.109).

\begin{figure}[ht]
\centering
\includegraphics[width=1.0\textwidth]{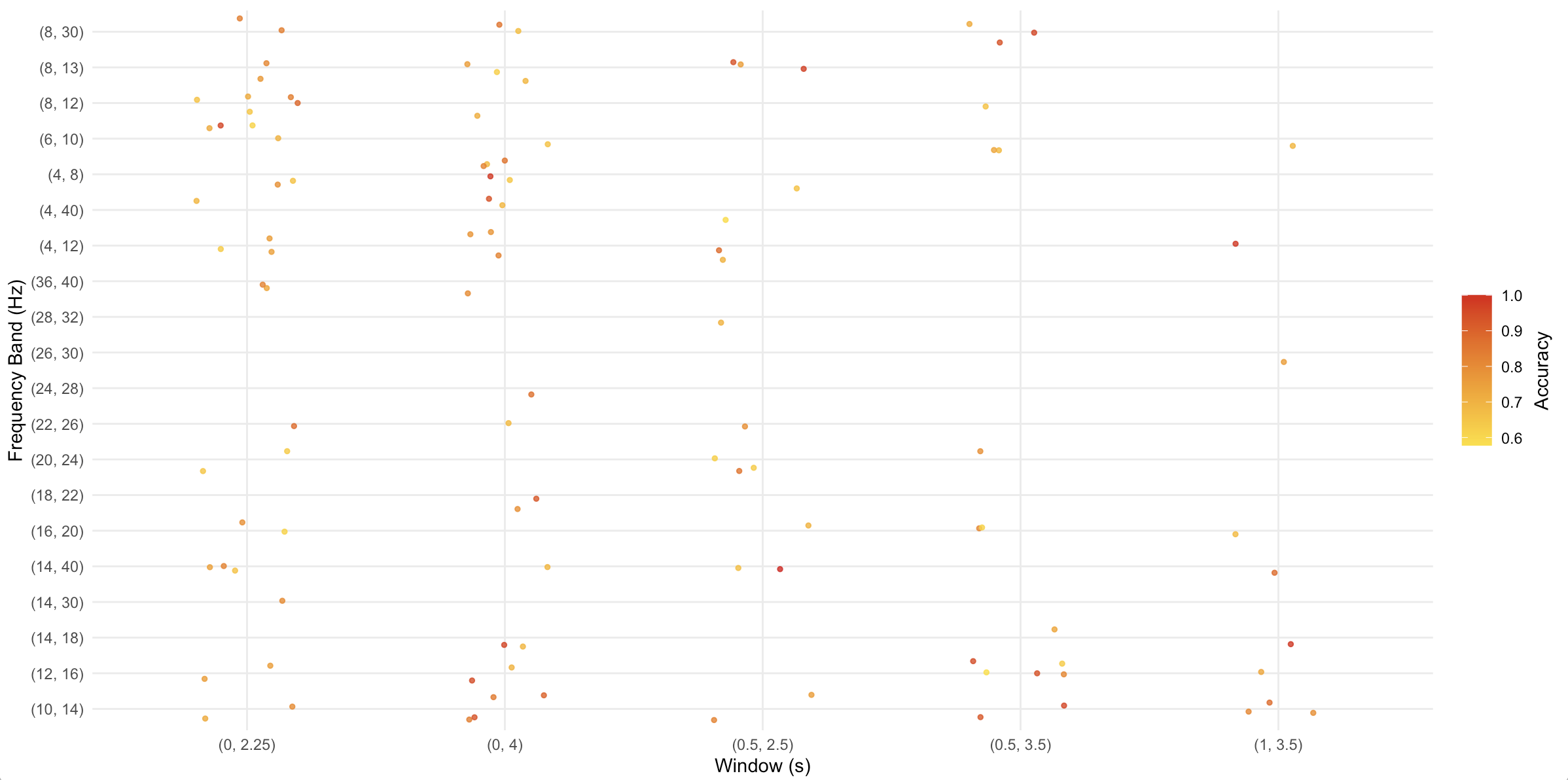}
\caption{\textit{MI EEG classification model accuracies obtained from the optimal combinations of chosen time windows (x-axis) and frequency bandwidths (y-axis) for each subject. The accuracies follow scaling from a heatmap (red denoting higher accuracies and yellow denoting lower accuracies).}}
\label{fig:fig2}
\end{figure}

The frequency band ranges occurring most prevalently (N >= 10) for the optimal time window and frequency band combinations are for (10, 14) Hz (N = 13, M = 0.844, SD = 0.071) and (12, 16) Hz (N = 10, M = 0.779, SD = 0.129). Here, the highest average classification accuracies were achieved for frequency bands of (14, 18) Hz (N = 4, M = 0.839, SD = 0.149), (4, 12) Hz (N = 9, M = 0.785, SD = 0.1), (4, 8) Hz (N = 8, M = 0.759, SD = 0.122) and (14, 40) Hz (N = 7, M = 0.779, SD = 0.128), where all frequency band ranges recorded a maximum accuracy of $0.97\dot{7}$, except for the latter which recorded a maximum accuracy of 1.

\section{Discussion}

This study investigated the contrast of classification performance of MI EEG models, trained on the PhysioNet EEG dataset, with different parameter combinations of time window and frequency band. The classification accuracies across 23 frequency bands during five different time windows demonstrated an optimal temporal and spectral combination of (0, 4) s at the range of (4, 12) Hz to yield the highest model accuracy on average across the participant population.
 
Use of cortical activity for at least 3 seconds after cue presentation for participant pre-MI baseline measures have been previously encouraged in the field of MI \cite{rimbert2023embc}. This study suggests the wider time window segment of 0 to 4 seconds and shorter time window segment of 0 to 2.5 seconds both effectively capture cortical activity changes elicited by MI. As participants may naturally perform mental imagery strategies at varying speeds upon cue presentation, accounting for the challenges of time latency during MI trials increases motor intent detection and classification accuracy \cite{Miao2021}.

While both alpha/mu and beta bandwidths are instead commonly opted for in MI-BCI protocols \cite{Meng2023}; \cite{rimbert2023embc}, our results point towards MI classification models performing more accurately using spectral data from the theta-alpha/mu bandwidth range. On the one hand, this could be explained by the broader frequency band range capturing more inter-subject variability within the alpha/mu oscillations \cite{Lotte2014}. On the other hand, this phenomenon may be indicative of MI attentional demands \cite{Tidare2021} as enhanced resting-state theta rhythms in the frontal areas prior to performing MI have also been previously associated with increased MI-BCI task performance \cite{Kang2021}.

The subject-specific ML approach is able to capture user variance and decrease the amount of redundant information, unlike subject-independent models. The importance of user-centred design is highlighted by the authors in \cite{Riascos2024}, who refer to the user as being "the core of the BCI interaction loop". Mental imagery classification models tailored specifically for its users may help to tackle the BCI inefficiency problem \cite{thompson2019critiquing}. Although the current paper reports the deduced optimal parameter combination across all participants, the ML models identified a large variety of optimal temporal and spectral parameter combinations for each participant at an individual level. While highlighting the natural differences in MI-induced cortical activity across subjects, these findings simultaneously underscore potential pitfalls of over-personalising MI EEG classification models. As user MI performance may fluctuate over time due to underlying cognitive and neuropsychological factors \cite{allison2010could}, such feature classification models should be adaptable for changing optimal parameter combinations on a participant-level. 

One of the most significant potential limitations of the personalised approach is the limited data sample size per participant, also serving as a practical limitation in the current study as seen by the low average accuracy scores for different combinations. There is a positive correlation between the increase in sample size and the increase in data classification performance. With roughly 90 trials per participant across three two-minute runs, each model is limited in its training data and therefore its potential for achieving higher accuracy values. This can be addressed through longer experimental runs, synthetic data generation or the implementation of a generalised approach, as subject-independent BCIs and addressing the model applicability challenge \cite{Lotte2014}. 

A subject-independent model would open up the possibility of implementing deep learning models due to the larger cumulative data pool from the combined subject population. A deep learning approach could allow for the capturing of complex, non-linear patterns \cite{sengupta2020review}, that linear models such as LDA fail to capture \cite{xu2025nimo}, at the sacrifice of personalisation. The use of generative AI and diffusion models to create synthetic EEG data for individual participants is another potential route for meeting the data size requirements of deep learning but without abandoning a subject-specific approach \cite{tosato2023eeg}.

Additional information on participant demographics such as age, sex, handedness or clinical history are not publicly available for our chosen dataset. This prevents the study from grouping participants using different characteristics and revealing relevant subpopulations within the cohort. Predicting motor function from within these clusters may be a useful way of increasing sample size without sacrificing model personalisation completely. 

Lastly, the current study’s model classification accuracy should be validated across multiple datasets and different classifier types, e.g. SVM or Multi-Layer Perceptron, to make more comprehensive conclusions of suggested parameter combinations.

\section{Conclusion}
This study investigated the impact of optimising both frequency band and time window options simultaneously on the classification performance of MI EEG tasks at a subject-specific level. An established MNE pipeline was modified to grid search through several parameter combinations of frequency band and time window for every subject, creating a unique LDA model for each combination. The results of the ML pipeline depicted the accuracy for each parameter combination per participant. 

Overall, there is a significant difference between both specific time window and specific frequency band options in terms of accuracy. The results suggest a combination of 0 - 4 second time window and a frequency band range of 4 - 12 Hz to increase model classification accuracy across subjects on average. Despite this, participants were often observed to obtain similar accuracies for other parameter combinations. This further highlights the need for a personalised approach accommodated to each MI BCI user’s characteristics.

In terms of future work, there is scope for a comparative study that contrasts the performance of the personalised approach to this problem against a generalised approach. By creating a classification model for each time window and bandwidth combination trained on all subjects' data, the optimal frequency bandwidth and time window combination to use on a more general population could be explored. 

One avenue for potential investigation could be the implementation of both deep learning and transfer learning in order to exploit the benefits of both generalisable and personalised approaches. Early layers in the neural network trained on a large cohort of participants could be frozen, with later layers being fine-tuned to patient-specific EEG patterns. In addition to this, the trade-off between the personalised approach (high specificity but little sample data) and the generalised approach (large amounts of data and low specificity) could be investigated. 

\begin{credits}
\subsubsection{\ackname}
This work was supported by the following institutions: Taighde Éireann - Research Ireland under grant number GOIPG/2025/7933, the Insight Centre for Data Analytics (12/RC/2289\_P2), CRT ML labs under Grant No.\ 18/CRT/6183 and by the Arizona State University\allowbreak{} \&\allowbreak{} Dublin City University Collaborative Doctoral Funding Programme, supported through DCU Biodesign Europe.
\end{credits}
%
%
%

\clearpage
\bibliographystyle{splncs04}     
\bibliography{bib}

\clearpage
\section*{Supplementary Figures}
\addcontentsline{toc}{section}{Supplementary Figures} 
\begin{figure}[ht]
    \centering
    \includegraphics[width=1\textwidth]{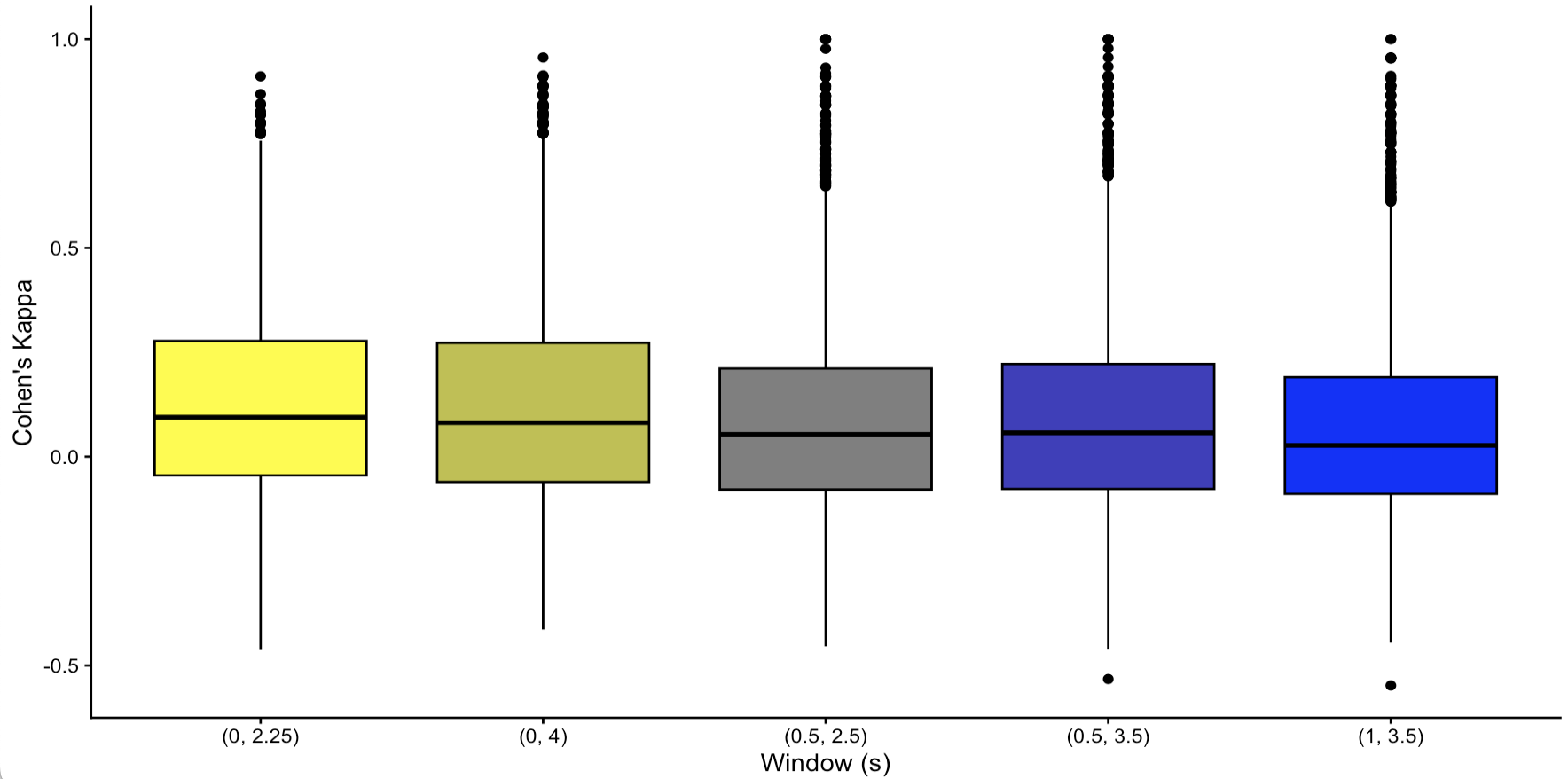}
    \caption*{\textit{Figure 4.2 MI EEG classification model Cohen’s Kappa obtained from varying time windows (seconds after cue).}}
\end{figure}

\begin{figure}[ht]
\centering
\includegraphics[width=1\textwidth]{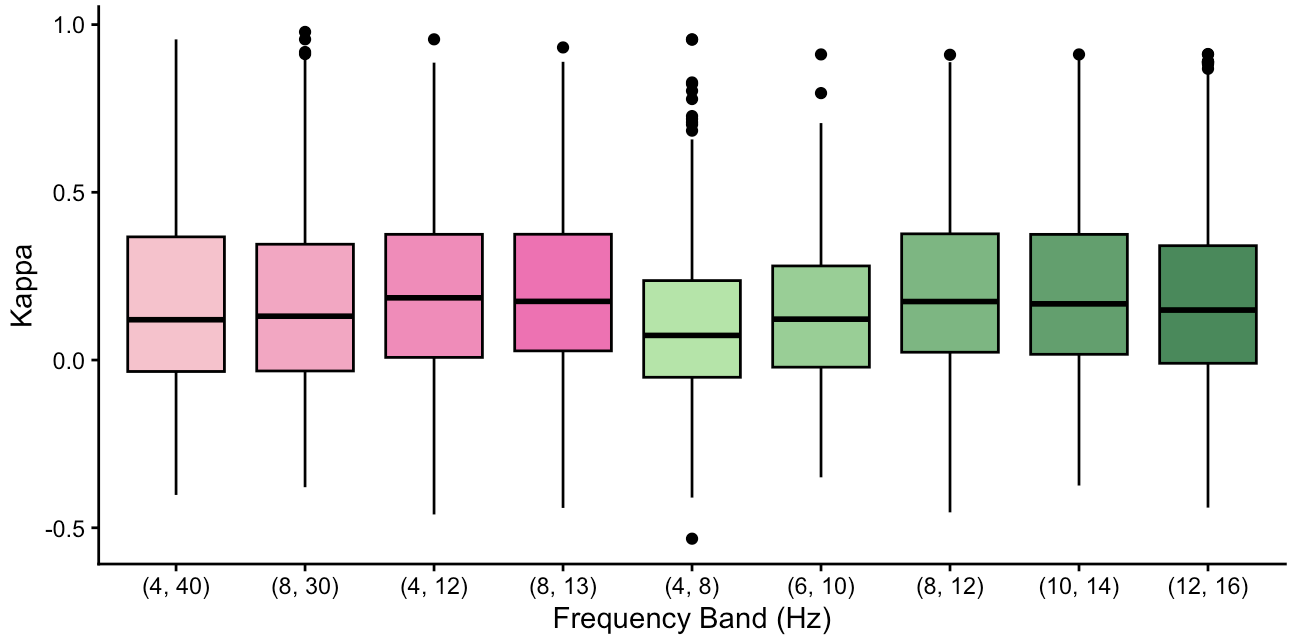}
\caption*{\textit{Figure 5.2 MI EEG classification model Cohen’s Kappa obtained from varying frequency bandwidths across combinations of broad bandwidths and alpha-specific bandwidths. The color differentiates between broad bandwidth (pink gradient) and specific bandwidths (green gradient).}}
\label{fig:supp2}
\end{figure}

\begin{figure}[ht]
\centering
\includegraphics[width=1.1\textwidth]{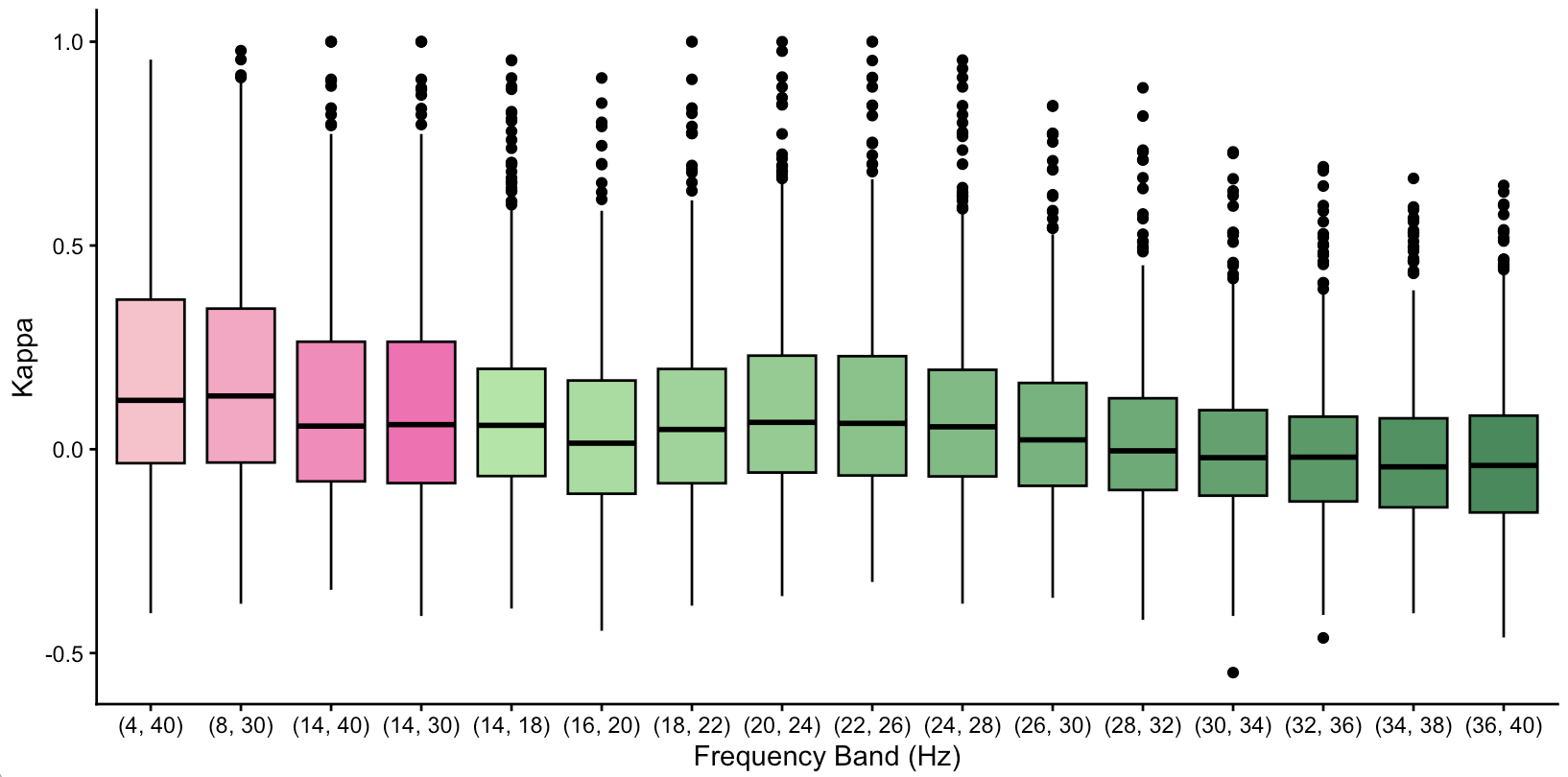}
\caption*{\textit{Figure 5.3 MI EEG classification model Cohen’s Kappa obtained from varying frequency bandwidths across combinations of broad bandwidths and beta-specific bandwidths. The color differentiates between broad bandwidth (pink gradient) and specific bandwidths (green gradient).}}
\label{fig:supp3}
\end{figure}

\begin{figure}[ht]
\centering
\includegraphics[width=1.1\textwidth]{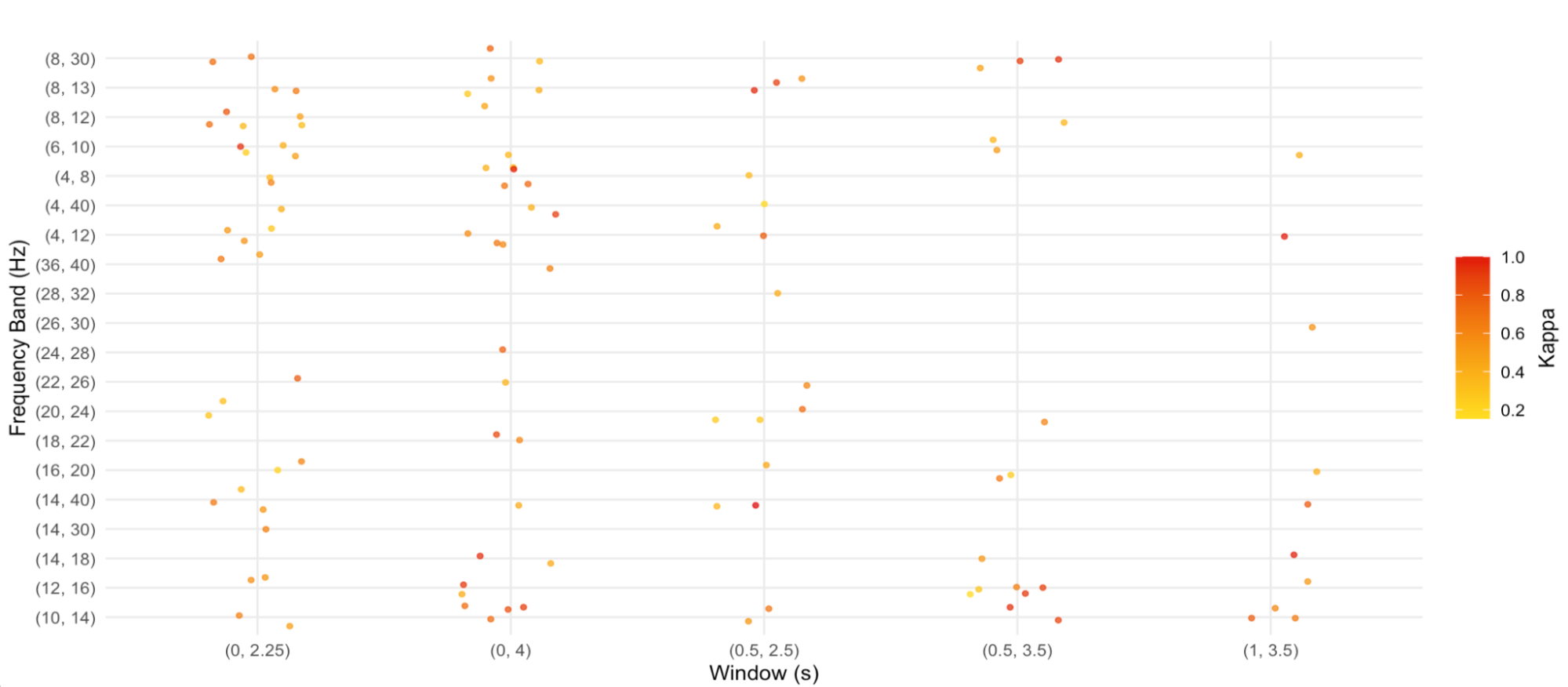}
\caption*{\textit{Figure 6.2 MI EEG classification model Cohen’s Kappa obtained from combinations of chosen time windows and frequency bandwidths.}}
\label{fig:supp4}
\end{figure}
\end{document}